\documentclass[%
 reprint,%
 amssymb, amsmath,%
cha,%
]{revtex4-1}

\usepackage{graphicx}

\begin{document}

\title{Measuring Political Polarization: Twitter shows the two sides of Venezuela}

\author{A. J. Morales, J. Borondo, J. C. Losada and R. M. Benito \\
Grupo de Sistemas Complejos. Universidad Polit\'ecnica de Madrid.  \\
ETSI Agr\'onomos, 28040, Madrid, Spain
}

\begin{abstract}
We say that a population is perfectly polarized when divided in two groups of the same size and opposite opinions. In this paper, we propose a methodology to study and measure the emergence of polarization from social interactions. We begin by proposing a model to estimate opinions in which a minority of influential individuals propagate their opinions through a social network. The result of the model is an opinion probability density function. Next, we propose an index to quantify the extent to which the resulting distribution is polarized. Finally, we apply the proposed methodology to a Twitter conversation about the late Venezuelan president, Hugo Ch\'avez, finding a good agreement between our results and offline data. Hence, we show that our methodology can detect different degrees of polarization, depending on the structure of the network.
\end{abstract}

\maketitle


\section{Introduction}

From a sociological point of view, polarization is a social phenomenon that appears when individuals align their beliefs in extreme and conflicting positions, with few individuals holding neutral or moderate opinions \cite{isenberg1986group,sunstein2002law}. 
Thus, as a process it is the increase of such divergence over time when people evaluate issues of diverse nature 
\cite{Baldassarri2007,Baldassarri2008,Dandekara2013}, like politics or religion. In
words of John Turner:
'Like polarized molecules, group members become even more aligned in the direction they were already tending'\cite{turner1987}.

In this paper, we propose a methodology
 to study the emergence of political polarization and quantify its effects. To this end, we introduce a model
 to estimate opinions, and a polarization index 
 that quantifies to which extent the resulting distribution of opinions is polarized.
 We say that a population is perfectly polarized when divided in two groups of the same size
and with opposite opinions. Hence, our measure of polarization is inspired by the electric dipole moment - a measure of the 
charge system's overall polarity. For two opposed point charges, the electric dipole moment
increases with the distance between the charges. Analogously, the polarization of two equally populated groups depends on how distant their views are.

As Downs argued in 1957 \cite{downs1957economic}, political discussion among individuals
minimizes the cost of becoming politically informed.  In other words, sensible individuals tend to
rely in the
opinions of experts instead of analyzing information by their own. In fact, several observational studies support this theory and suggest that the expertise distribution within a social network affects the political communication patterns\cite{huckfeldt2001social}. 
Hence, by controlling the opinion of a minority of influential individuals and mapping
the communication fluxes among the population we can estimate their distribution of opinions. To this end, we propose a model based on DeGroot model \cite{degroot1974reaching}. The original model proposed by DeGroot describes how a group of individuals might reach a shared opinion, by iteratively updating their opinion as the average of their current opinion with the opinions of their neighbors. 
Such global coordination, without centralized control, can also be efficiently achieved when individuals adopt the majority state of their neighbors, even in the presence of noise or complex topologies \cite{Amaral2004}. Recently, the DeGroot model has been
used to study the conditions under which consensus is achieved \cite{acemoglu2011opinion,golub2010naive,jackson2010social}. 
However, as consensus is rarely reached in real world \cite{Krackhardt09,benczik00}, variants of this model can held 
to a diversity of opinions \cite{bindel2011bad,acemoglu2013opinion,krause2000discrete,friedkin1990social}. 

In contrast to opinion generation models, such as the voter model \cite{CLIFFORD01121973,Holley75,Suchecki2005}, we do not aim to study the evolution of opinions, but to infer a distribution of opinions formed on a social network from which to measure polarization.
In our model, a minority
of influential individuals propagate their opinions through a directed network influencing the remaining individuals. Thus,
each individual iteratively updates her opinion according to her incoming neighbors-those influencing her. Hence, by taking advantage of complex network analysis \cite{newman_rev}, we are able to 
estimate the opinion of the whole majority that a priori was unknown. 
The behavior of the influential minority is similar to zealots in the voter model \cite{Mobilia2003,Mobilia2007}, but their impact
in the model's dynamics is different. In our model, zealots, rather than preventing consensus, allow us to infer the opinions of all the nodes in the network. Contrary to the voter model where opinions are binary (0 or 1), the opinions in our model represent a continuous distribution.
 In absence of polarization, the expected resulting distribution of opinions would be a 
narrow distribution centered at a neutral opinion. However, as
 polarization emerges, the resulting distribution shifts to a bimodal distribution  with two peaks emerging around the two 
 dominant and confronted opinions \cite{Dixit2007}.

How can political polarization be detected and therefore be fixed? Nowadays, digital traces of human collective behavior
 \cite{Lazer2009} represent an opportunity to detect and measure in real time different phenomena, such as polarization. In fact 
 political segregation
has already been observed on political blogs \cite{Adamic2005} or Twitter \cite{Conover2011,Borondo2012}. Recent research has shown that the most prominent and politically active users mainly interact with their own partisans \cite{Adamic2005,Conover2011,Borondo2012}, leaving little 
space for real debate and cross ideological interactions. 
However, segregation does not necessarily imply polarization, as two separated groups of people that share the same opinion can not be considered as polarized. Hence, in order for a population to be polarized, the opinions of the two groups should also be conflicting or opposed\cite{guerra2013measure}.  In the latter part of this paper we show how to
 apply our methodology to online data gathered from Twitter in order to estimate 
individuals opinions and to measure the emergent political polarization. Twitter
provides an interesting context in which to study polarization as it represents a wide
variety of different types of communications, going from personal to those coming from traditional mass media. In this platform, a
 minority of $elite$ users concentrate much of the collective attention, but still a big fraction of the content they produce reaches 
 the mass through intermediaries or 'opinion leaders' \cite{burt1999social}. In other words, the 'two-step-flow' of communication
is still valid on Twitter \cite{wu2011says}.

We begin this paper by proposing a model to estimate opinions in which a minority of influential individuals propagate their opinion through a social network influencing the opinions of the remaining individuals. Thus, the result of the model
is a probability density function $p(X)$, that determines the fraction of individuals holding an opinion $X$. Next, we introduce
the polarization index to measure the political polarization from the resulting opinion distribution. To 
illustrate the power of the methodology, we apply it to a Twitter conversation regarding the death announcement of the 
Venezuelan President (Hugo Ch\'avez). Finally, we contrast the results with offline data.
 
\section{Estimating Opinions}
\label{opinion}

 \begin{figure*}
 \begin{center}
 \includegraphics[width=.8\paperwidth]{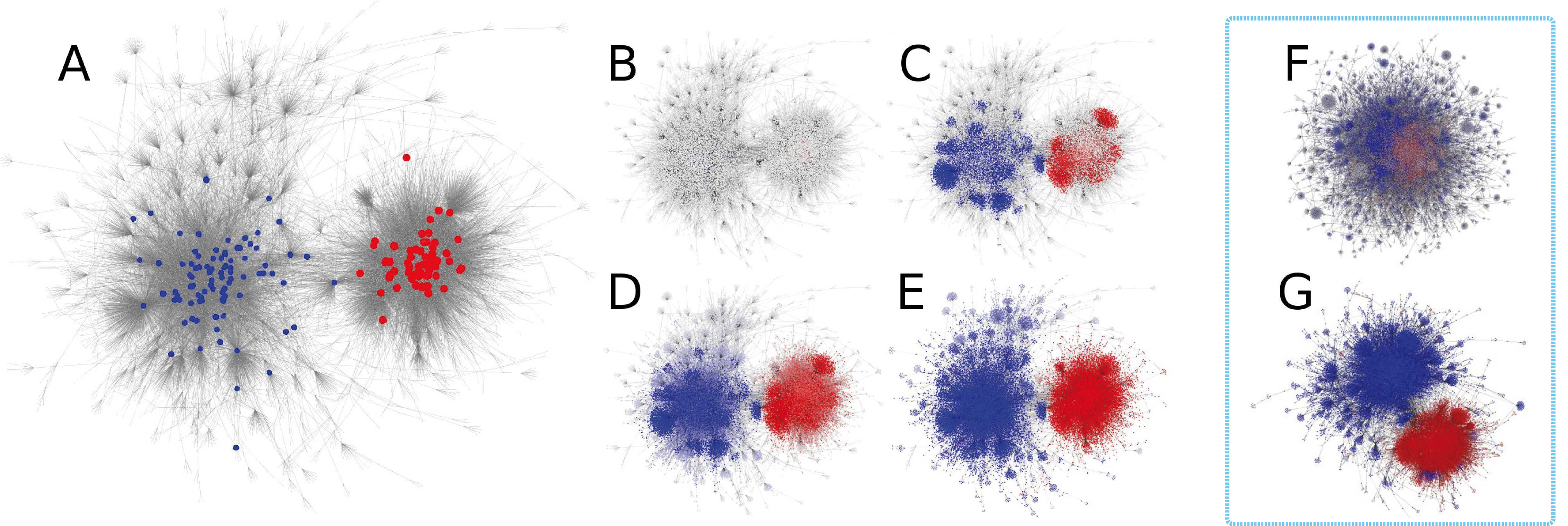}
 \caption{Schema of the influence spreading process in the opinion estimation model. (A) Displays the seed nodes in the network, colored according to their respective ideology. (B) Displays the network at $t=0$, before seeds start to propagate their influence. (C) Shows the state of the network at $t=1$. (D) shows the state of the network at $t=n/2$. (E) Displays the final state of the network at $t=n$.  (F) and (G) Visualizations of two examples of the result of the opinion formation model to the Venezuelan dataset for non polarized (F) and polarized (G) days.} \label{WuHu}
 \end{center}
 \end{figure*}
 
We present a model to estimate the opinions of individuals who interact on a social
network, in order to obtain their opinions distribution. In it we distinguish two types of individuals, $elite$ and $listeners$.
The first ones have a fixed opinion and act like seeds of influence, while the opinion 
of the second ones depends on their social interactions. The model is fully specified by the following assumptions:

1. \textbf{Initial Conditions:} The world is abstracted by a directed network, $G$, in which each individual is represented by a node and links account for influence rather than friendship or other kind of relationship. We define two different  subset of nodes, $S$ accounting for  $elite$; and $L$, accounting for  $listeners$. Additionally we endow each $elite$ with a parameter, $X_s$, that determines her opinion value and that will remain constant for the duration of the model. $X_s$ lies in the range, $-1 \leq X_s\leq 1$, where 1 and -1 represent the two extreme and confronted poles. Finally we set an initially neutral opinion, $X_l(0)=0$ to all $listeners$.

2. \textbf{Opinion Generation:} At each iteration, $elite$ nodes, $S$, propagate their opinions
through the established network, $G$,
influencing $listeners$, $L$. Hence, each listener iteratively updates her opinion value as the mean opinion value of her incoming neighbors. Thus the opinion at time step, $t$, of a given listener, $i$,
is given by the following expression:
 
 \begin{equation}
 X_i(t)=\frac{\sum_j{A_{ij}X_j(t-1)}}{k_i^{in}}
  \label{eq_X}
 \end{equation}
 
 where $A_{ij}$ represents the elements of the network adjacency matrix, which is 1 if and only if there is a link from $j$ to $i$, and $k_i^{in}$ corresponds to her indegree. The process is repeated until all nodes converge to their respective $X_i$ value, lying in the range $-1 \leq X_i \leq 1$.
Thus, the results of the model are given in a density distribution of nodes' opinion values $p(X)$. Note that
the opinions of individuals do not depend on their opinion in the previous step. This is because we are
estimating their opinion that a priori was unknown, rather than studying the evolution of opinions.

The dynamics of the model is illustrated in Fig. \ref{WuHu}, where we present an schema of the influence spreading process. Panel A visualizes the instantiation of the model where each $elite$
node has been 
colored according to her opinion (red, $X_s=-1$; and blue, $X_s=+1$). Panels B-E show
the dynamics of the influence process from the initialization (B) to the final converged state (E). Panels
(F) and (G) visualize two empirical networks corresponding to a non polarized (F) and a polarized (G) case.

\section{Introducing a new measure of polarization in opinion distributions: the polarization index}
\label{Indicators}

\begin{figure}
 \begin{center}
 \includegraphics[width=250px]{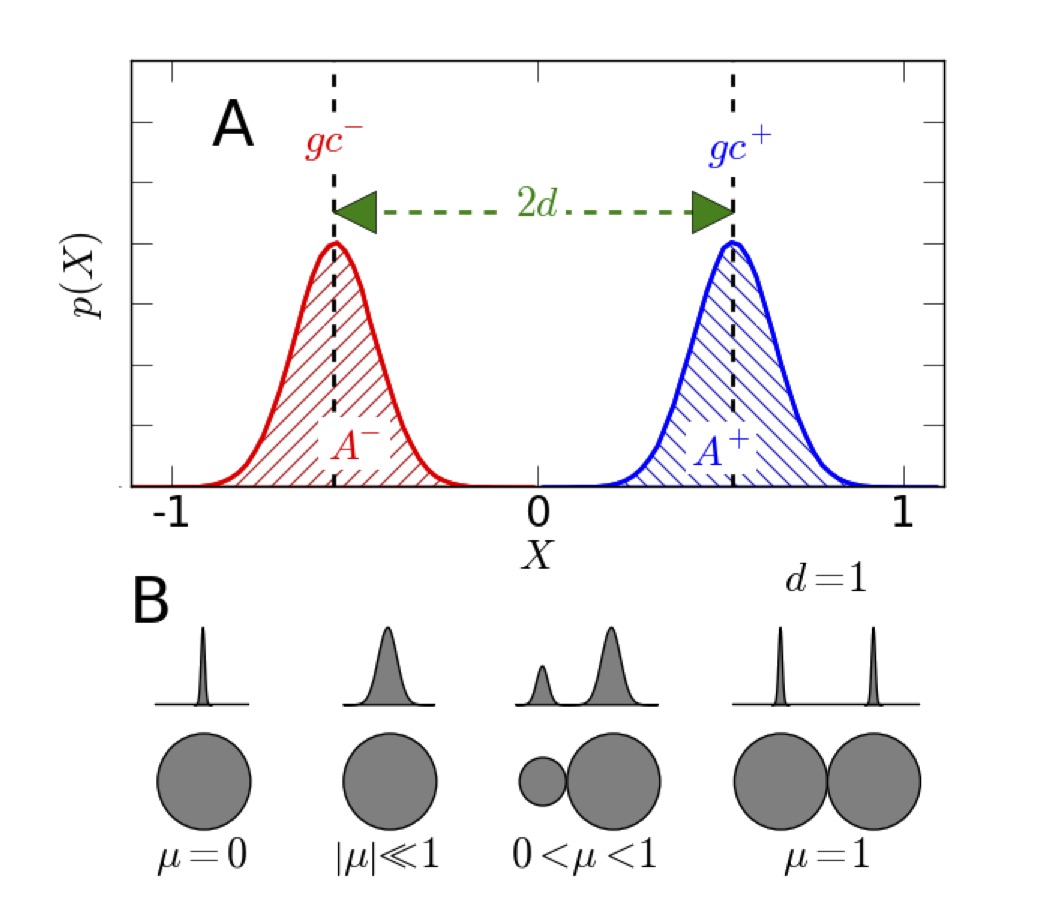}
 \caption{Schema explaining polarization and the proposed index $\mu$. (A) Density distribution of  opinions. $gc$ stands for the gravity center of each pole, $A$ stands for the area associated to each ideology, and $d$ stands for the pole distance. (B) Visualization of the polarization index, $\mu$,  given in eq. \ref{eqpol}, for four situations.} 
 \label{casos}
 \end{center}
 \end{figure}

We say that a population is perfectly polarized when divided in two groups of the same size
and with opposite opinions. Hence, we propose a measure of polarization that quantifies both effects
for the resulting $X$ distribution obtained from our model. This definition is
inspired by the electric dipole moment- a measure of the charge system's overall polarity. In the 
simplest case of two point charges of opposite signs ($-q$ and $+q$) the electric dipole moment
is proportional to the distance among the charges. This is analogous to a simple
scenario consisting of two persons with different ideologies, thus the polarization depends on how
conflicting their points of view are ({\em i.e.} the distance among the two ideologies).

We begin by calculating the population associated with each opinion (positive and negative).
To this end, we define $A^-$ as the relative population of the negative opinions ($X < 0$). By the same
token, we define $A^+$ as the relative population of the positive opinions ($X > 0$). Hence, both 
variables can be expressed as:

\begin{eqnarray}
 A^- = \int^{0}_{-1}p(X) dX=P(X<0) \; ,\\  
A^+ = \int^{1}_{0}p(X) dX=P(X>0)
 \end{eqnarray}

So we can express the normalized difference in population sizes, $\Delta A$ , as:

\begin{equation}
\label{deltaA}
 \Delta A = |A^+ - A^- |= | P(X>0) - P(X<0)|
 \end{equation}

Next, we quantify the distance between the positive and negative opinions. In other words we measure how differing the opinions of the two sides are. To this end we determine the gravity center of the positive and negative  opinions that can be written as:

 \begin{eqnarray}
 gc^- = \frac{\int^{0}_{-1}p(X)X dX}{\int^{0}_{-1}p(X) dX}  \; , \\
 gc^+ = \frac{\int^{1}_{0}p(X)X dX}{\int^{1}_{0}p(X) dX}
 \end{eqnarray}

and define the pole distance, $d$,  as the normalized distance between the two gravity centers. Hence, it can be expressed as:
 
 \begin{equation}
 d = \frac{|gc^+-gc^-|}{|X_{max}-X_{min}|} = \frac{|gc^+-gc^-|}{2}
 \label{distance}
 \end{equation}

This formula gives $d=0$  when there is no separation between the gravity centers, {\em i.e.} there are no longer two differentiated groups and everyone shares a similar opinion; and $d=1$  when the two 
opinions are extreme and perfectly opposed.

Finally, we can use eqs. \ref{deltaA} and \ref{distance} to write down a general formula to
measure polarization as a function of the difference in size between both populations $\Delta A$
 and the poles distance $d$. Thus, we define the {\it polarization index}, $\mu$, as:
 \begin{equation}
 \mu = (1-\Delta A)d
 \label{eqpol}
 \end{equation}
 
 This formula gives $\mu=1$ when the distribution is perfectly polarized. In this case the opinion distribution function is
 two Dirac delta centered at $-1$ and $+1$ respectively. Conversely, $\mu = 0$ means that the opinions are not polarized at all, and the resulting distribution of opinions would either take the form of a single Dirac delta centered at a neutral opinion, or be entirely centered in one of the poles, implying that the population (A) of the other pole would be reduced to zero and $\Delta A=1$. Notice that for non-uniform distributions centered in a neutral opinion, $|\mu| \ll 1$, but still presents a minimum polarization due to a small separation between gravity centers, that depends on the standard deviation $\sigma$. In the case of a Gaussian distribution centered at zero, $\mu = \sigma \sqrt{2/\pi}$.

In between,
 polarization can lie within the range, $0<\mu<1$, for three reasons: i) The population sizes associated to each opinion are equal,
 but the pole distance $d$  is lower than $1$. ii)  Despite $d$ being equal to $1$, the population sizes associated to each opinion are different 
 and therefore there is a majority sharing a similar opinion. iii) A combination of i and ii. Fig. \ref{casos}A illustrates the basic concepts of the proposed index
 of polarization, as it visualizes the area associated to 
each opinion, their corresponding gravity centers and the pole distance for a standard case of a perfect bimodal distribution. 
In panel B of this figure, we have visualized  non polarized distributions ($\mu =0$ and $|\mu| \ll 1$), a perfectly polarized one ($\mu=1$) and a case in between.

 \begin{figure}
 \begin{center}
 \includegraphics[width=3in]{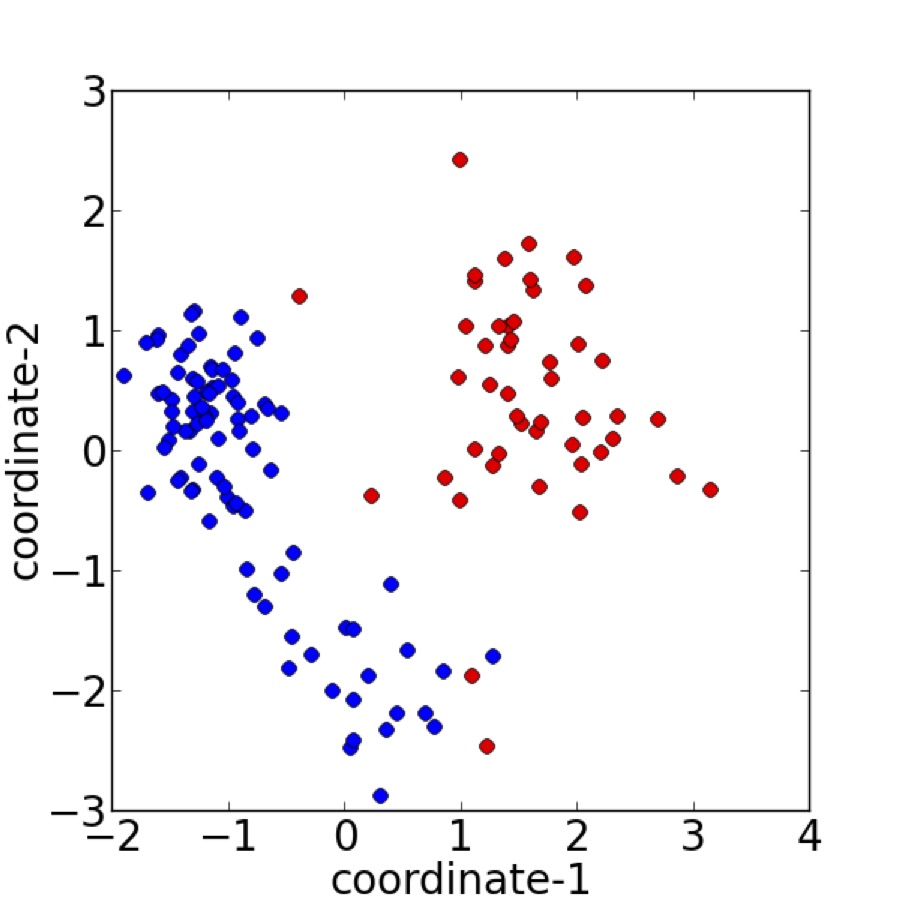}
 \caption{Projection in a two-dimensional space of the distribution of $elite$ users according to the similarity of their content. Dots represent users and colors indicate the community they belong to in the $elite$ network: red for the officialism and blue for the opposition. The distance between users is inversely proportional to the similarity of their content.} \label{texpol}
 \end{center}
 \end{figure}

 \begin{figure*}
 \begin{center}
 \includegraphics[width=500px]{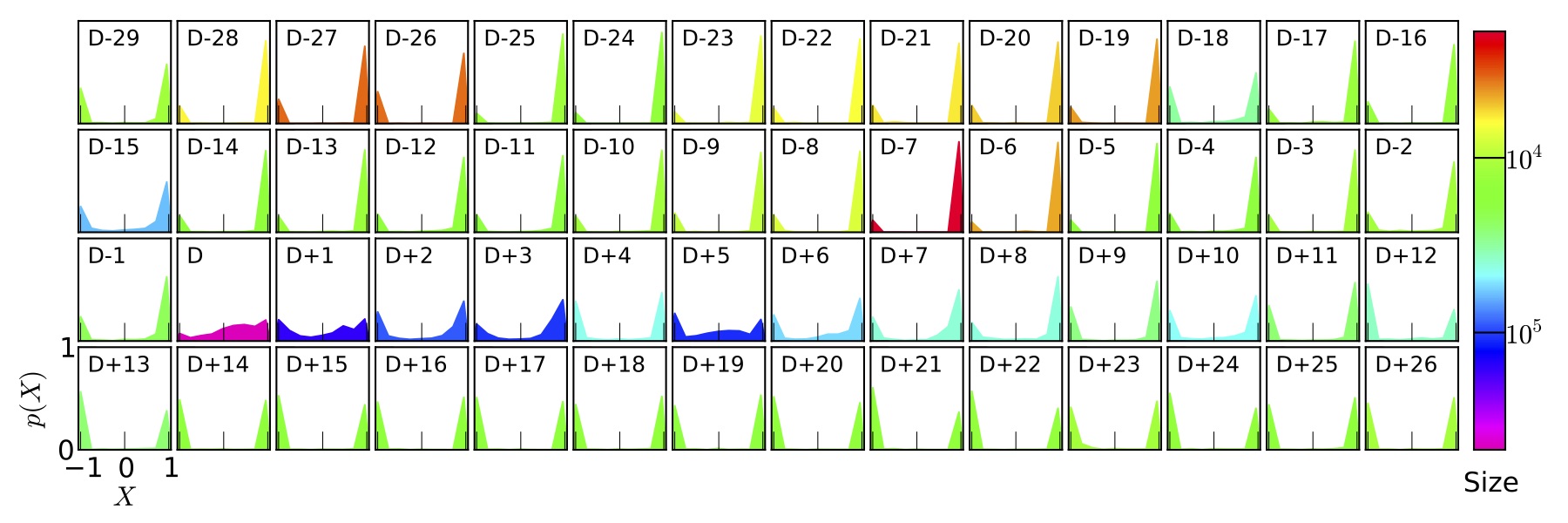}
 \caption{Time evolution of ideological value ($X_i$) probability density functions ($p(X)$) for the Venezuelan conversation. Labels indicate the day of observation, $D$ standing for the day of the Presidents death. Colors indicate the number of participants.} \label{sabana}
 \end{center}
 \end{figure*}
 
\section{Twitter data: The Venezuelan case}
\label{Ven}

In this section, we apply our model and polarization index to Twitter data regarding the late Venezuelan President Hugo Ch\'avez. We downloaded over 16,383,490 messages written by 3,173,090 users from 02/04/2013 to 05/04/2013. This period covers one month preceding his death, the announcement of the death, and the schedule for new elections. We use retweets
as a proxy for influence \cite{Boyd2010,Dann2010,Jaebong2012,Stewart2013,Liere2010,Shan2011,Weber2014}, and build a weighted and directed network accounting for the adoption of ideas among Twitter users for each day. Whenever a user $i$ retweets a message originally posted by  user $j$, we assume that $i$ is being influenced by $j$'s ideas.
 Hence, a new directed link ($j \rightarrow i$) is created. We constructed an individual retweet network for each day of the observation period, which is a total of 56 networks. More details 
about the dataset and the retweet networks can be found in the Appendix A and B respectively.

In order to apply the model to these daily networks, we begin by defining a set of $elite$ users. We denote as $elite$ those users who gained a noticeable amount of retweets and actively participated in the conversation along the observation period. The distribution of users according to the total amount of retweets obtained ($S_{out}$) and participation rate ($\rho$) is shown in Fig. \ref{ParticipationDay} of Appendix B.
In this case, we considered a very small set (0.02\%) of influential users who participated most of the observation period ($\rho > 89\%$) and obtained a very high number of retransmissions ($S_{out}> 1000$).

The $elite$ users mainly correspond to politicians, journalists and mass media accounts, whose political position and editorial tendency are publicly known and who belong to both sides of the Venezuelan political spectrum. In order to assign them an ideology value, $X_s$, we first studied their network of interactions. In the $elite$ network, nodes represent the $elite$ users, and links are created and accumulated whenever an $elite$ user $i$ retweets an $elite$ user $j$. This network is polarized in a well defined two-community structure, with modularity $Q = 0.38$. In each community, users share political ideology and hardly interact with users from the other pole. In fact, the assortative mixing \cite{Newman2003} by political ideology is very high ($r=0.88$). 

In order to further understand the $elite$ polarization, we analyzed the content of their messages. For this purpose, we abstracted each $elite$ user as a high-dimensional vector, where each element represents the number of times that the user posted each of the 500 mostly used words from all the $elite$'s messages. Then, we reduced the high-dimensional space into a two-dimensional one, by applying a multi-dimensional scaling algorithm \cite{MDS}. In this algorithm, users are mapped into a new space by preserving the distance between them in the original one. This means that the distance between users is inversely proportional to the similarity of their posted contents. In Fig. \ref{texpol}, we present the projection of the users in the new two-dimensional space. Dots represent users and colors are assigned according to the community they belong to in the $elite$ network. It can be noticed that these users are not homogeneously distributed in the new space. Instead, they are separated from each other in agreement with our previous classification. This means that the use of language is polarized among the $elite$ users.

\begin{figure}
 \begin{center}   
 \includegraphics[width=250px]{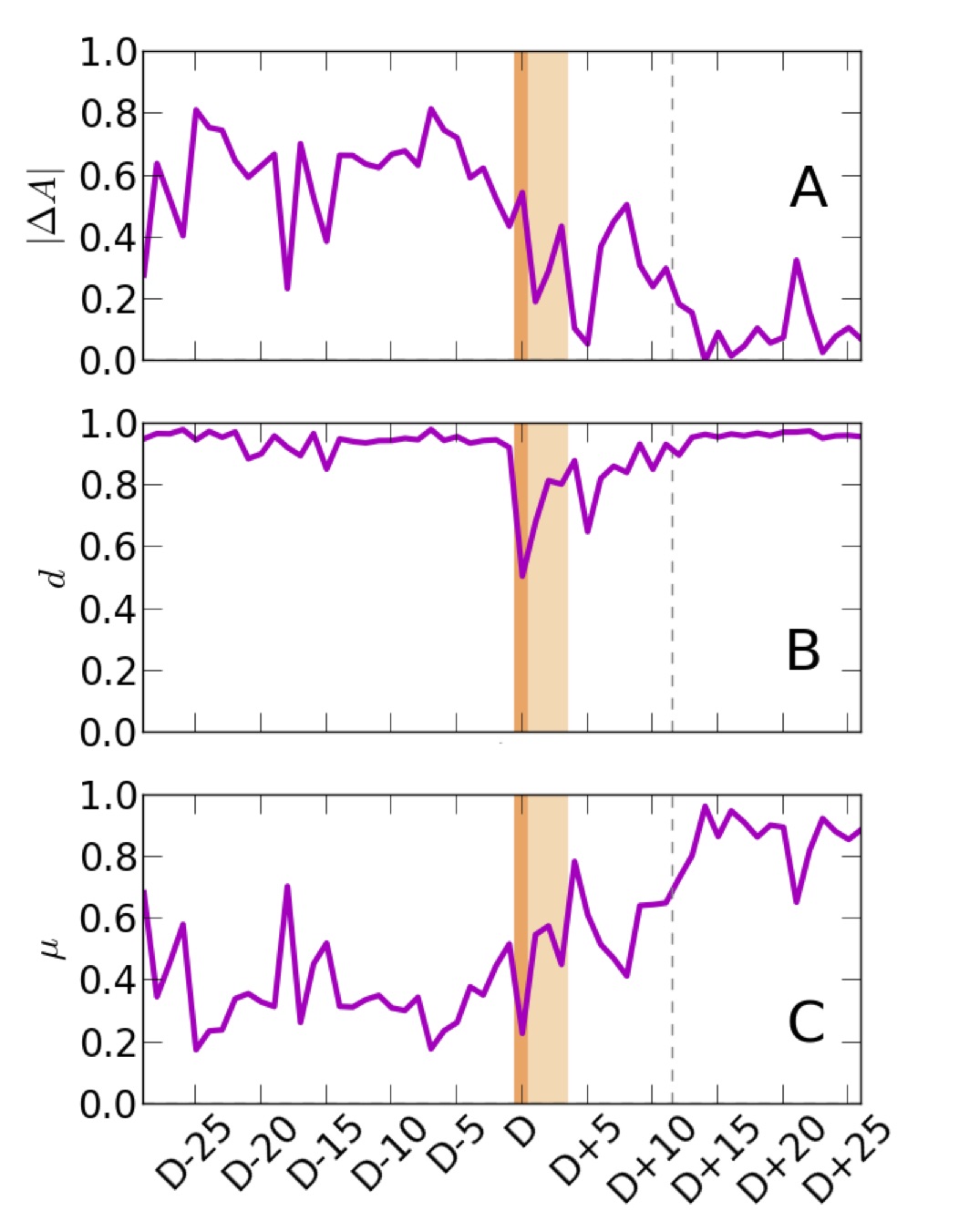}
 \caption{Time evolution of the polarization index $\mu$ (C), and the variables associated with it: difference in population sizes (A) and pole distance $d$ (B) for the Venezuelan conversation.} \label{pol}
 \end{center}
 \end{figure}

After identifying the $elite$ users, we assigned them ideology values of $X_s=-1$ to the officialism side and of $X_s=1$ to the opposition. The remaining users ($99.98\%$) were assigned the role of $listeners$ and $X_l = 0$. After running the model we obtained an ideology probability density function $p(X)$ for each day. The resulting $p(X)$ for each network are presented in Fig. \ref{sabana}. The label indicates the day of observation, $D$ representing the day of the death. The color indicates the network size in terms of the number of participants. As can be
seen the days with largest participation (purple and blue) correspond to the most important
announcements: the presidents death (day $D$), and call for election (day $D+6$). Next, we calculated the 
polarization index ($\mu$), pole distance ($d$) and populations sizes for the resulting distributions of each day and plotted the results in Fig. \ref{pol}.

We identify day $D$ as a turning point which ended up polarizing even more the conversation.
During the days preceding the announcement (from $D-29$ to $D-1$), $X$ presents a bimodal
distribution in which the officialism population (negative side of the $X$ distribution) is considerably smaller than the opposition 
(positive side of the $X$ distribution). This means that during this period the conversation was still polarized, but practically 
monopolized by the opposition. Hence, despite the fact that the pole distance reached values over $0.9$, the polarization index just 
averaged under $0.4$. Then a shift in the conversation emergent patterns took place on the day of the President's death 
announcement (day $D$). During this day $X$ lost its bimodal distribution, and the resulting $p(X)$ was centered
 around neutral values, minimizing  the pole distance. All these meaning that the conversation was not so polarized and that the network does not have a two-island structure anymore. Therefore, the 
 polarization index decreased, $\mu\approx0.25$. 
This behavior is due to the bursty growth of the conversation at day $D$ (see Fig. \ref{size} in the Appendix B). As a consequence, the previously segregated modules combined into a single-island structure, many times larger than the usual network size. 
 Besides a large amount of users from all around the
 globe joined to the conversation, making the topic international, rather than local from Venezuela. In fact, during this day the 
 percentage of users tweeting from Venezuela ($\approx 20\%$) was very low in comparison to the rest of the days (average around 
 $>80\%$). Hence, our set of Venezuelan $elite$ were not capable of polarizing this majority of worldwide users.
  However, from there on the conversation recovered its bimodal distribution of opinions. Moreover, the polarization reached its
  maximum from day $D+12$ (marked with the dashed line) onwards, day that the officialism new leader entered the 
  conversation. From this day onwards  $X$ presents a bimodal distribution, where 
 the populations of both sides are similar. Therefore, the polarization index 
averaged values around $0.9$.

\section{Twitter shows the two sides of Venezuela}
\label{GeoPol}

\begin{figure}
 \begin{center}
 \includegraphics[width=250px]{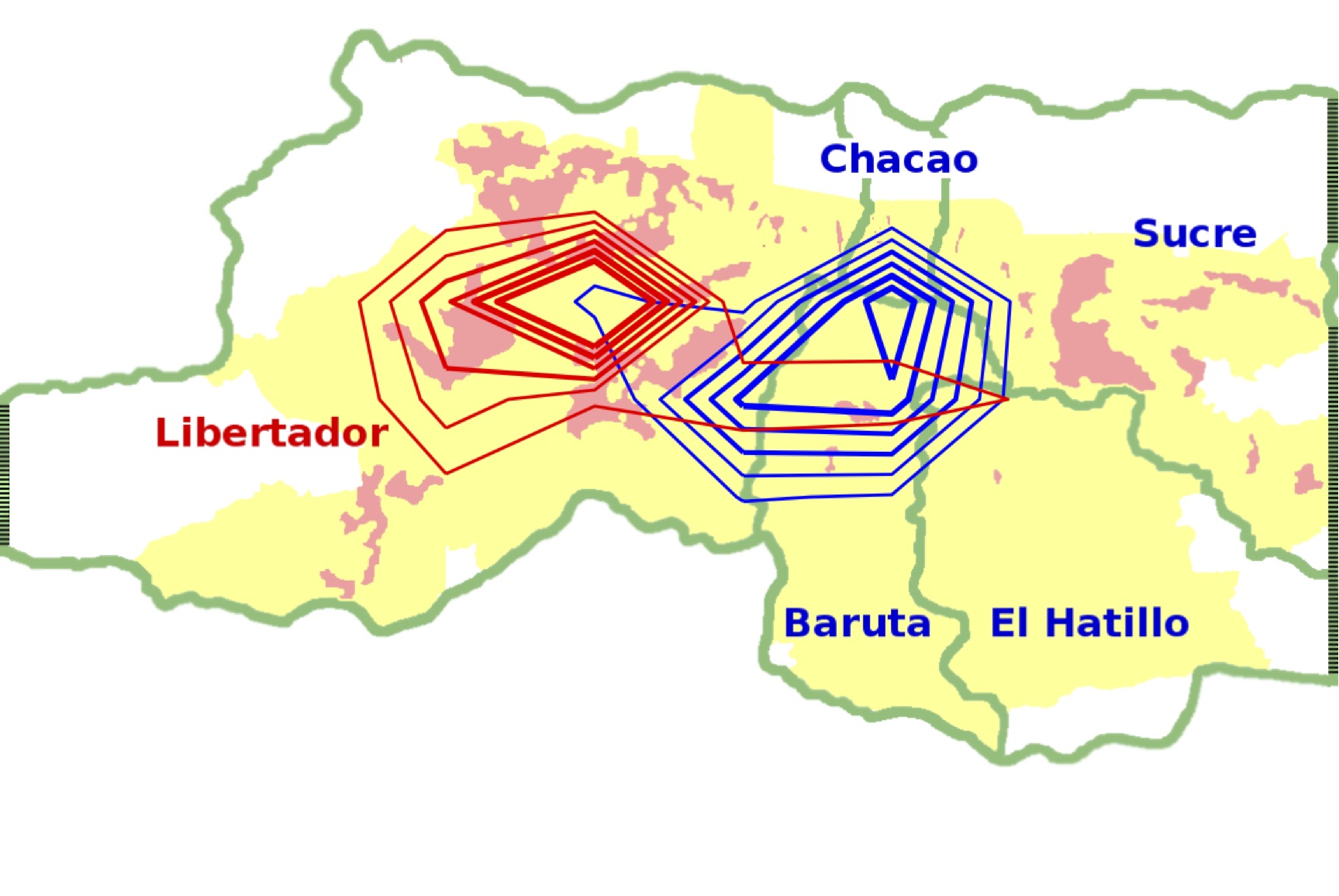}
 \caption{Geographical polarization in the city of Caracas. Contour lines represent the density functions of the probability that a tweet associated with the officialism (red) or the opposition (blue) had been posted by a geolocated user at a given position (latitude and longitude). These contours have been superimposed to the map of Caracas, Venezuela. From inside out, contours indicate the following values: [0.175, 0.15, 0.0125 0.10, 0.075, 0.05]. The green lines border the five municipalities composing the city. Labels indicate the name of the municipality and the color indicate the ruling party according to the 2013 Venezuelan local elections (red for the officialism party and blue for the opposition parties). White represents unpopulated areas, yellow urbanized areas and pink the poorer neighborhoods.} \label{ven}
 \end{center}
 \end{figure}

Next we evaluate our model and the validity of Twitter data by comparing the geographic distribution of the polarized users with offline data regarding the Venezuelan socioeconomic and political landscape. More specifically, we analyze  the geographical density of geolocated tweets in Caracas, the capital city of Venezuela, taking the results obtained from the most
polarized days in section \ref{Ven} as a proxy of their ideology. For this purpose, we have built the density functions that a tweet associated with the officialism or the opposition had been posted by a geolocated user at a given position (longitude and latitude). We considered a grid of 100 cells between longitudes [-67.12$^o$, -66.71$^o$] and latitudes [10.31$^o$, 10.57$^o$] and counted the number of tweets in each cell, identified with each ideology. Then, we normalized both counts by their respective total number of tweets. The resulting functions are two surfaces on top of the map, which we show in Fig. \ref{ven} as contour plots (red for the officialism and blue for the opposition) that indicate lines of equal value in the 2-D probability density function. These contour lines are superimposed on a map of the municipalities composing the city of Caracas. There are five of them, bordered in green. The labels correspond to the municipality name, and the color indicates the ruling party-like the officialism in Libertador and the opposition in Chacao, Sucre, Baruta and El Hatillo. Additionally, urbanized areas are colored in yellow and 
poorer regions (slums) in pink. Notice that the West region is characterized for having lower income and governed by the officialism, while the East part is wealthier and governed by the opposition. 


It can be noticed that the regions where each pole concentrates most of their tweets are well separated from each other, showing that the city presents a clear geographical polarization. In fact, there is a good correspondence between the results of our model and offline evidence,
such as electoral results or socioeconomic factors. Those municipalities governed by the opposition contain the highest concentration of users identified with this pole, and the same effect occurs for the officialism side of the political spectrum. We also have to remark that the areas with higher concentration of users aligned with the officialism, correspond to the parts of the city with the largest concentration of poorer neighborhoods (pink areas). Conversely, the opposition users concentrate in urban developed regions. All these suggesting that the basis of the Venezuelan popular polarization resides in socioeconomic factors and that the political conflict in Venezuela presents a strong territorial facet. 

\section{Conclusions}

 Modern democracies have to represent the conflicts existing in our society, while at the same time maintain the 
 social stability \cite{Diamond1990}. However, as polarization emerges, the few most powerful parties tend to capitalize the 
 whole of the public attention and support, silencing the moderate opinions and under representing minorities. Consequently, todays' society is concerned about polarization, as a politically polarized society
implies several risks. These risks include the appearance of radicalism or
civil wars. In fact, one of the actual challenges
and a cutting edge topic is how to detect the emergence of political polarization and how
to fix it. 

We state that the possibility to gather user generated data from social media platforms \cite{Lazer2009}, together with network science \cite{linked}, represents an opportunity to detect political polarization. In this work, we have proposed a methodology to study and measure the emergence of polarization from social interactions. We have used it, to analyze the political polarization in one of the most polarized countries: Venezuela \cite{Ellner2004,Morales2012}.  We have done this, by applying our methods to a Twitter conversation 
about the late Venezuelan president Hugo Ch\'avez. We have shown that our methodology is able to detect different degrees of polarization in the conversation, depending on the participants' behavior, given by the structure of the network. Finally, we have contrasted our results against offline data,
such as municipality governments or socioeconomic factors, finding a good correlation between
the online and offline polarization. Hence, we conclude that online data seem to be a good proxy to detect politically polarized societies, as the online polarization that we found is a reflection of the Venezuelan political, territorial and social polarization.
 
Another relevant question is: Can social media platforms help reduce political polarization
as more voices could be heard? 
Although we do not answer this question, our results show that a minority of $elite$ users were able to influence the whole online social
network, resulting in a highly politically polarized conversation. However, these Venezuelan local 
influential accounts were not capable of polarizing the network when the conversation stopped being local of Venezuela and turned to be international. 
This opens two questions that can be studied
from a social media analysis perspective: i) How does online political polarization change at
different scales-like city, country, continent or whole world? ii) How could we target interventions in control strategies on social media that might be implemented to reduce polarization?

\begin{acknowledgments}
This research was supported by the Ministry of Economy and Competitiveness-Spain under Grant No. MTM2012-39101 
\end{acknowledgments}

\section*{Appendix A: Datasets}
\label{datsets}

In this work, we analyze messages from the online social network Twitter. We downloaded data from a temporal index of tweets managed by the Search API v1 \cite{twapi}, whose limitations are specified as the result of queries complexity and frequency, instead of fixed a percentage of the main stream. We queried for messages mentioning the name of the late Venezuelan President Hugo Ch\'avez, during the events that surrounded his disease and death in 2013. We considered a two month period from February 4th, 2013 (29 days before the death announcement) to April 4th, 2013 (26 days after the death announcement). In summary, we downloaded 16,383,490 messages posted by 3,173,090 users from more than 159 countries (according to the 0.4\% of geographically located messages). Our analyses are based on those messages that represent reweets (49\% of the downloaded content) and more specifically those that constitute the larger components of the communication networks, which were posted by 57\% of original set of users.

The Venezuelan Internet penetration represents about 40\% of the population, where most of users belong to middle and middle-low class \cite{tendenciasdigitales}. Online social networks are very popular in this country. Around 33\% of Venezuelans use Facebook \cite{tendenciasdigitales} and almost 10\% use Twitter  \cite{Semiocast2012}. In fact, Venezuela ranks thirteenth out of all countries in number of Twitter users \cite{Semiocast2012}. Moreover, Venezuela has the highest proportion of mobile Internet in Latin America at over 30\% of total connections, due to the popular use of social media from mobile phones \cite{gsmamobileeconomylatinamerica}. 

The political usage of Twitter in Venezuela is of great importance and has played a fundamental role in the recent Venezuelan history \cite{WSJ2012,Nagel2012}. The late President Hugo Ch\'avez was considered to be the second most influential world leader on Twitter \cite{DigitalDaya2012}, preceded only by the US President Barack Obama. The collective who opposes the late President, also finds on social media a channel to freely speak to their supporters and protest against the Government \cite{Morales2012}.


\section*{Appendix B: Networks}
\label{networks}

We have built one retweet network for each day of the observation period (56 networks). A retweet network emerges from user-to-user interactions during the message retransmission process provided by Twitter. Nodes represent users and links are created between users $i$ and $j$, when $i$ forwards the content previously posted by $j$. Edges are weighted in proportion to the frequency that $i$ retweeted $j$'s messages, and directed in the sense of the flow of information from the message source $j$ to the retweeter $i$.

\begin{figure}
\begin{center}
\includegraphics[width=3.2in]{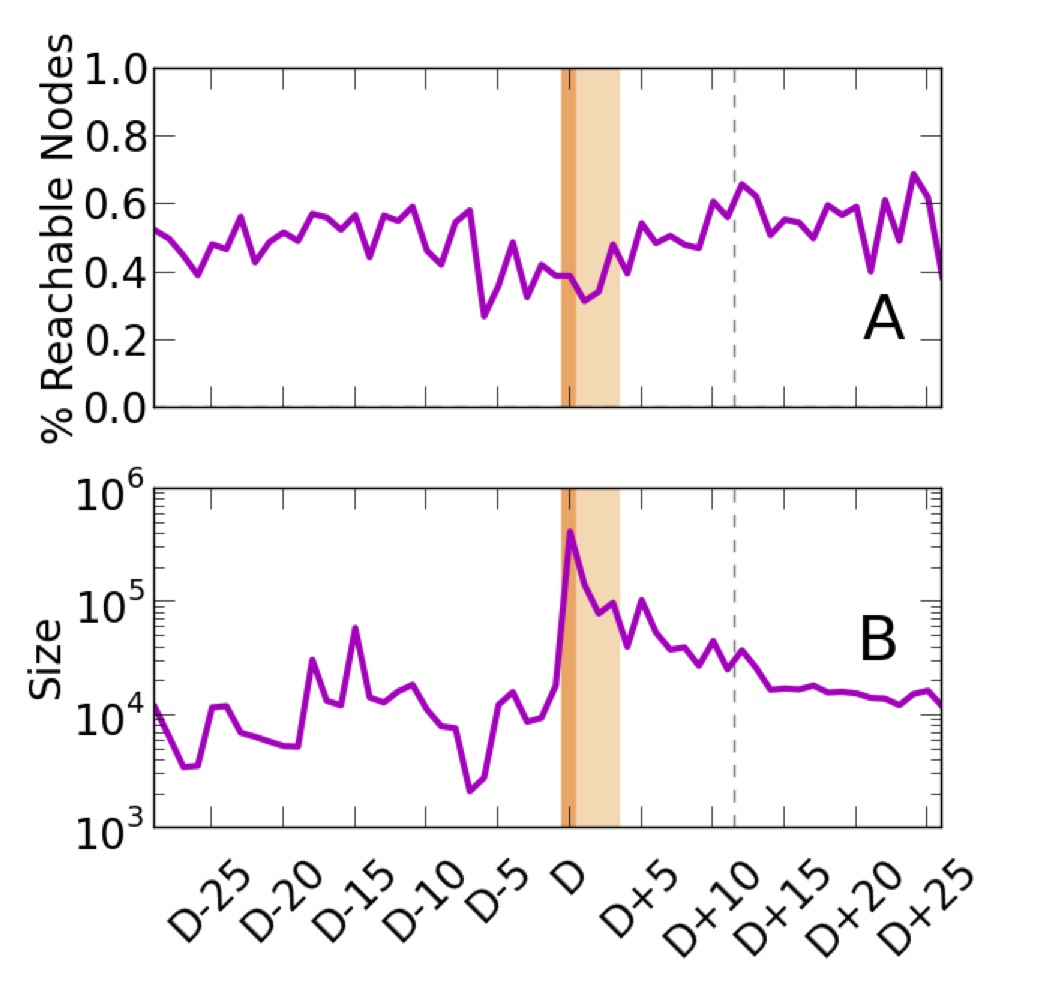}
\caption{Time evolution of the relative number of reachable nodes in comparison to the GC (A) and size of the reachable nodes' networks (B).}
\label{size}
\end{center}
\end{figure}

A single network contains several retransmission cascades, seeded and propagated by the conversation participants. When these cascades are aggregated, several disconnected network components emerge. Among these components, there is a single one called Giant Component (GC) whose size is in the same order of the whole network. As part of the GC, there is a set of nodes that are reachable from the set of influential {\it $elite$}, that represent about 50\% of the GC's size (Fig. \ref{size}A). For most of days, the amount of reachable nodes fluctuated around 10,000 users and explosively grew to almost 500,000 users during day $D$ (Fig. \ref{size}B). This behavior is typical of breaking news and critical events \cite{Yang2011,Bagrow2011}, with a bursty increase during the main occurrence and a slow decay that may last for several days. 

\begin{figure}
\begin{center}
\includegraphics[width=3.5in]{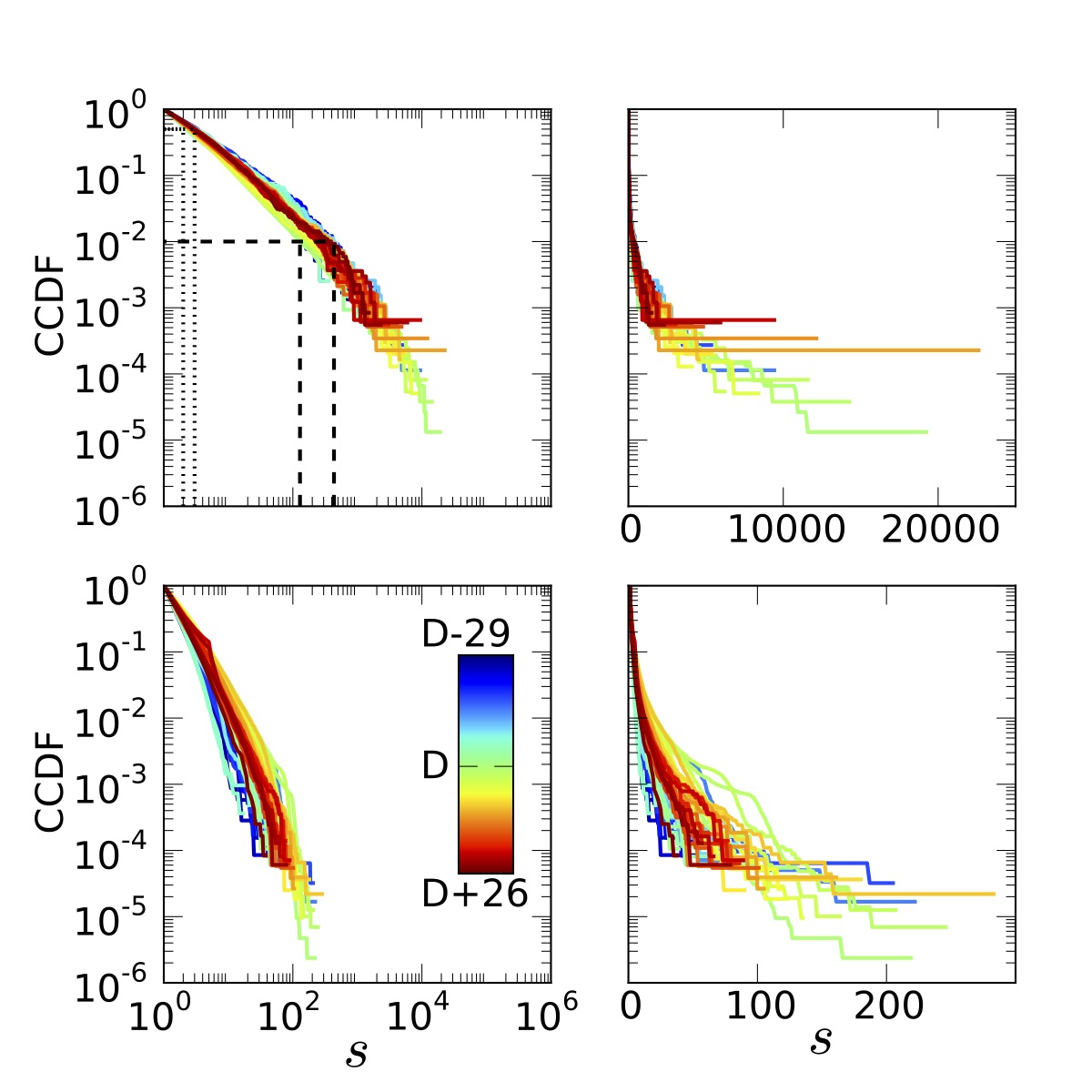}
\caption{Complementary cumulative density function (CCDF) of the retweet networks out strength $s_{out}$ (top) and in strength $s_{in}$ (bottom), from the Twitter conversation about the Venezuelan President Hugo Ch\'avez, in log-log (left) and linear-log (right) scale. The colors indicate the corresponding day of the observation period. The dotted line indicates the range of $s_{out}$ for 50\% of the population, while the dashed lines indicate the range of $s_{out}$ for 1\% of the population.}
\label{ComponentsGrados}
\end{center}
\end{figure}

The retweet networks characterize the way that the collective attention is organized during an event on Twitter. The out strength ($s_{out}$) indicates the amount of retweets gained by a participant, while the in strength ($s_{in}$) indicates the number of retweets made by the participant. In Fig. \ref{ComponentsGrados} we have superimposed the 
out strength (top) and in strength (bottom)
complementary cumulative density functions (CCDF) for each of the constructed networks, 
in log-log (left) and linear-log (right) scales.
In both cases, the distributions display heterogeneous behavior, being the out strength distributions broader than the in strength distributions.
In order to compare, whether these distributions behave like an exponential rather than a power law, we calculated the likelihood ratio statistical test \cite{Clauset2009,Alstott2014}. We found that the probability of these distributions to follow an exponential curve, instead of a power law, has a $p$-value $< 0.01$ for more than 98\% of the outgoing distributions and 75\% of the incoming distributions, where over 87\% of the distributions have a $p$-value $< 0.05$. 

From a dynamical point of view, the power law distributions imply a preferential attachment mechanism  \cite{linked}, where the chances of being retweeted increases with the number of retweets previously gained. These dynamics result in heterogeneous distributions where the great majority of users receive a very small amount of the collective attention, while some scarce users receive a disproportionally larger amount of it. For example, at all days 50\% of the population gained between 2 or 3 retweets at most (dotted lines in the top left panel of Fig. \ref{ComponentsGrados}), while the 1\% of most retweeted participants gained from 130 to 430 retweets as minimum (dashed lines in the top left panel of Fig. \ref{ComponentsGrados}).

To further understand the relationship between the individual activity and the attention received, we will aggregate the observation period by characterizing the individuals according to their rate of participation and total amount of retweets gained. The participation rate is defined as:

\begin{equation}
\rho = \rho_i/T
\end{equation}

where $\rho_i$ is the number of days that the user $i$ actively participated in the retweet process and $T$ is the total length of the observation period. The total number of retweets gained by user is measured as:

\begin{equation}
S_{out}= \sum_{t=0}^T{s_{out}(t)}
\end{equation}

where $s_{out}(t)$ is the out strength of the node $i$ at day $t$. If the user did not actively participate at day $t$, then $s_{out}(t) = 0$.

\begin{figure}
\begin{center}
\includegraphics[width=3.5in]{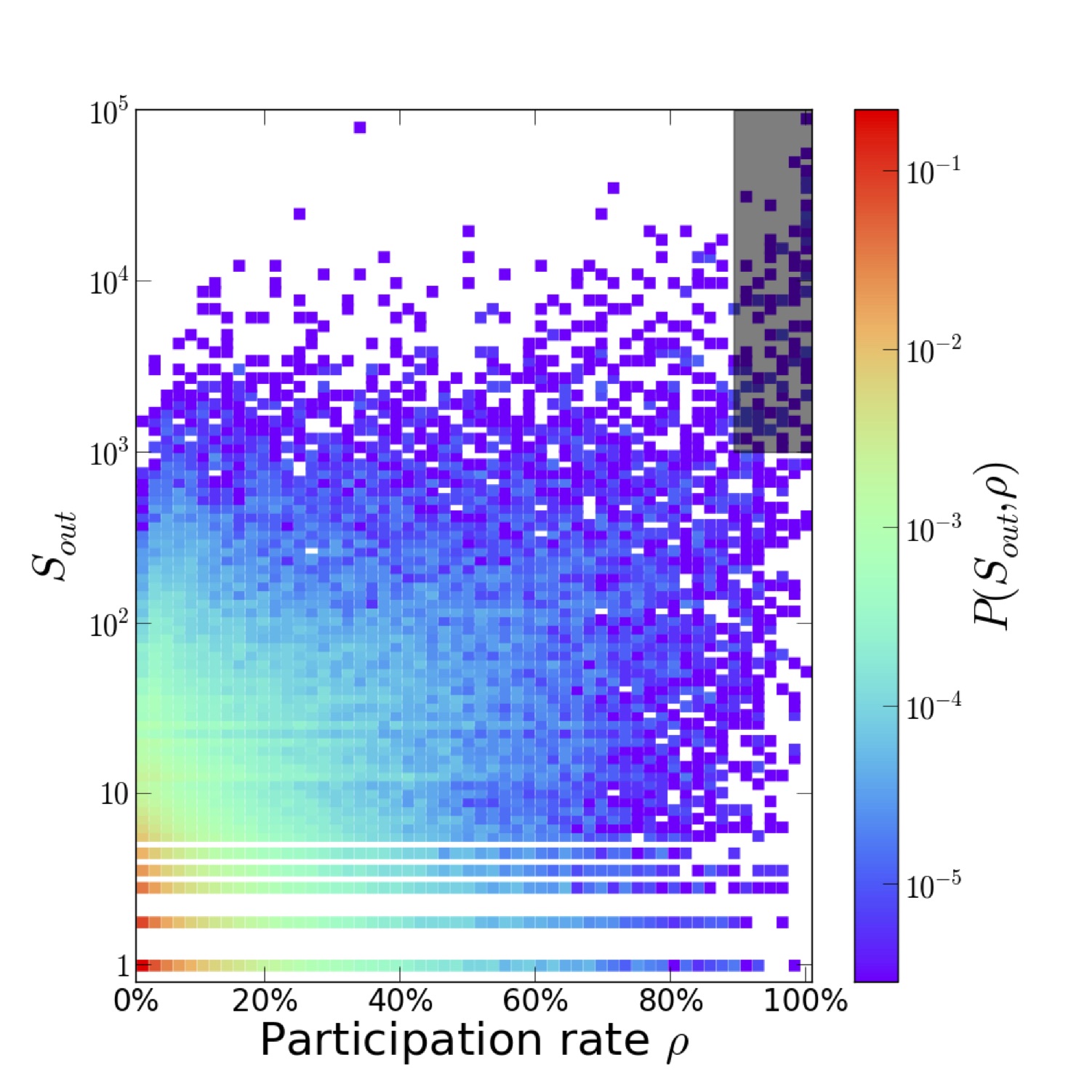}
\caption{Joint probability density function of the accumulated out strength ($S_{out}$) and the participation rate ($\rho$), from the Twitter conversation about the Venezuelan President Hugo Ch\'avez. Colors correspond to the density of users. The grey reactangle marked at the top right corner indicates the $elite$ users defined in section \ref{Ven}.} 
\label{ParticipationDay}
\end{center}
\end{figure}

The joint probability density function of the accumulated out strength $S_{out}$ and the participation rate $\rho$, $P(S_{out},\rho)$, is shown in Fig. \ref{ParticipationDay}. This distribution indicates the total amount of attention received by users according to their participation rate. It can be noticed that the largest density of users (red and orange dots in Fig. \ref{ParticipationDay}) participated less than 20\% ($\rho < 0.2$) of the days and present a small out strength value ($S_{out} < 10$), which means that most of them received a little amount of the collective attention. However, there is a very small set of users at the upper right corner in Fig. \ref{ParticipationDay}, who participated almost every day and present an extremely high $S_{out}$. This minority of highly influential users captured most of the collective attention throughout the observation period, and define the  {\it $elite$} users considered in section \ref{Ven}.

\end{document}